\documentclass[twocolumn]{aastex62}
\usepackage[mediumspace,mediumqspace,Grey,squaren]{SIunits}

\graphicspath{{./}{figures/}}

\submitjournal{ApJ}

\shorttitle{Multiple populations in NGC 2121}
\shortauthors{Li \& de Grijs}

\begin{document}

\title{When does the onset of multiple stellar populations in star
  clusters occur? Detection of enriched stellar populations in NGC
  2121}

\correspondingauthor{Chengyuan Li}
\email{chengyuan.li@mq.edu.au}

\author{Chengyuan Li} 
\affil{Department of Physics and Astronomy, Macquarie University,
  Sydney, NSW 2109, Australia}
\affil{Centre for Astronomy, Astrophysics and Astrophotonics,
  Macquarie University, Sydney, NSW 2109, Australia}
\affiliation{Key Laboratory for Optical Astronomy, National
Astronomical Observatories, Chinese Academy of Sciences, 20A Datun
Road}
  

\author{Richard de Grijs}
\affiliation{Department of Physics and Astronomy, Macquarie
 University, Sydney, NSW 2109, Australia}
\affil{Centre for Astronomy, Astrophysics and Astrophotonics,
  Macquarie University, Sydney, NSW 2109, Australia}
\affiliation{International Space Science Institute--Beijing, 1
 Nanertiao, Zhongguancun, Hai Dian District, Beijing 100190, China}

\begin{abstract}
Star-to-star light-element abundance variations, know as multiple
stellar populations (MPs), are common in almost all Galactic globular
clusters. Recently, MPs have also been detected in a number of massive
clusters with ages in excess of 2 Gyr in the Large Magellanic Cloud
(LMC), thus indicating that age is likely a control parameter for the
presence of MPs. However, to conclusively confirm this notion,
additional studies of star clusters in the LMC's `age gap' of 3--6 Gyr
are required. Here, we use {\sl Hubble Space Telescope} observations
to study the 3 Gyr-old cluster NGC 2121. Compared with so-called
`simple' stellar population models, the cluster's red-giant branch
exhibits an apparent spread in a specific color index that is
sensitive to intrinsic chemical spreads. { The observed spread can
  be explained by an intrinsic spread in nitrogen abundance of
  $\sim$0.5--1.0 dex}. NGC 2121 has a comparable mass to its younger
counterparts without MPs, indicating that cluster mass might not be
the dominant parameter controlling the presence of MPs. The transition
phase between the occurrence of clusters with or without MPs seems to
occur at ages of 2--3 Gyr, indicating that cluster age may play a
dominant role in the establishment of MPs.
\end{abstract}

\keywords{globular clusters: individual: NGC 2121 --
  Hertzsprung-Russell and C-M diagrams}

\section{Introduction} \label{S1}

The notion that all star clusters are simple stellar populations
(SSPs), with color--magnitude diagrams (CMDs) described by a single
isochrone of fixed age and metallicity, is a view of the past.
Current observations show that almost all Galactic globular clusters
(GCs) are composed of multiple stellar populations (MPs), exhibiting
multiple subgiant branches \citep[SGBs;][]{Piot12a}, red-giant
branches \citep[RGBs;][]{Piot15a}, main sequences
\citep[MSs;][]{Piot07a}, and sometimes combinations of these features.
Spectroscopic studies have demonstrated conclusively that such
multiple features can be explained by star-to-star chemical
variations, including spreads in C, N, O, Na, Mg, Al, and in some
cases iron content \citep{Cann98a,Carr09a,Mari09a,Panc17a}. Although
common in GCs, variations in the light elements are rarely found in
the field or in open clusters \citep[e.g.,][]{Macl15a}. Sometimes,
chemically enriched field stars also have kinematics that are similar
to those of GCs \citep{Tang19a}. These results all lead to the
suggestion that GCs represent the only known environment that appears
capable of producing enriched stellar populations.

Various models for the origin of MPs have been proposed
\citep[e.g.,][]{Meyn06a,Decr07a,Derc08a,Deni14a,Bekk17a}. Most draw on
self-pollution of the intra-cluster gas during the early stages of
cluster evolution, and they all identify massive single or binary
stars as the main polluters. All of these models imply that the total
cluster mass is the key parameter that controls the appearance and
significance of MPs: after all, the more massive the cluster is, the
deeper is its gravitational potential well capable of accreting the
polluted intra-cluster gas.

Although this correlation between the significance of the MPs and
cluster mass has been confirmed for old Galactic GCs \citep{Milo18a},
it is unclear if the fraction of the enriched population also
correlates with cluster mass in other galaxies. First, the fractions
of enriched stars in extragalactic clusters are highly uncertain
\citep{Nied17a}. In addition, to date only the stellar populations in
clusters of satellite galaxies \citep[e.g., in the Large and Small
  Magellanic Clouds (LMC and SMC) or the Fornax dwarf;][]{Mart18a}
have been resolved; most of these clusters are younger than the
Galactic GCs \citep{Lars12a,Dale16a,Mart18a}. Since clusters will lose
mass during their evolution, younger clusters with comparable masses
to those of the Galactic GCs will eventually evolve to become less
massive than most GCs by the time they reach similar ages. The 2.2
Gyr-old LMC cluster NGC 1978 is, thus far, the youngest cluster known
to exhibit MPs \citep{Mart18a}. Intriguingly, its slightly younger,
1--2 Gyr-old counterparts appear to be fully chemically homogeneous
\citep[e.g., NGC 1806;][]{Mucc14a}. These results suggest that age
might be another important factor controlling the appearance of MPs.

One popular hypothesis suggests that the observed MPs may have formed
through non-standard stellar evolutionary effects associated with
stellar rotation \citep[e.g.,][]{Bast18a}. Most clusters younger than
2 Gyr exhibit extended main-sequence turn-off regions (eMSTOs) in
their CMDs \citep[e.g.,][]{Milo09a,Cord18a}, which are likely caused
by differential stellar rotation \citep{Cord18a}. Note, however, that
MPs seem to occur at ages where the eMSTOs have already disappeared.
\cite{Mart18a} suggest that magnetic fields may play a role in
generating chemical anomalies. They suggest that MPs might be a
specific feature of low-mass stars featuring strong magnetic
fields. Such stars cannot be rapidly rotating because of magnetic
braking \citep{Card07a}. This hypothesis, although as yet speculative,
implies that older clusters without eMSTOs should exhibit MPs.

To date, NGC 1978 is the only known cluster exhibiting MPs that is
younger than 3 Gyr. The stellar populations of clusters with ages
between 3 and 6 Gyr are still poorly studied, however. To determine
the exact age of the onset of MPs, it is important to study clusters
with ages in the LMC's so-called `age gap' between 3 and 6 Gyr
\citep{Piat02a}. NGC 2121 is one of the few suitable clusters in this
context. The most recent dedicated study of NGC 2121 dates from 18
years ago \citep{Rich01a}. Its authors derived a cluster age of $3.2
\pm 0.5$ Gyr, with a metallicity of only one-quarter solar ([Fe/H] =
$-0.6 \pm 0.2$ dex).  In addition, unlike most intermediate-age LMC
clusters \citep[e.g.,][]{Milo09a}, NGC 2121 exhibits a sharp MSTO,
indicating that most of its member stars are non- or slowly rotating
stars, which in turn implies that most have been affected by magnetic
braking. NGC 2121 thus provides a unique case to (i) examine whether
the occurrence of MPs is a specific feature associated with low-mass
stars (since the masses of the RGB and MS turn-off stars in NGC 2121
have decreased to below the critical mass for magnetic braking to
cease being important), and (ii) constrain the age range associated
with the onset of MPs.

In this paper, we examine whether NGC 2121 hosts MPs among its RGB
population. Using ultraviolet (UV)--optical--infrared (IR)
observations obtained with the {\sl Hubble Space Telescope} ({\sl
  HST}), we will show that MPs are indeed present along the cluster's
RGB. This paper is organized as follows. In Section 2 we introduce our
data reduction. In Section 3 we summarize our analysis approach and
present our main results. We also compare our results with SSP
models. In Section 4 we present a discussion and our conclusions.

\section{Data Reduction} \label{S2}

We use observations obtained with both the {\sl HST}'s Ultraviolet and
Visual Channel of the Wide Field Camera 3 (UVIS/WFC3) and the Wide
Field and Planetary Camera (WFPC2). The UVIS/WFC3 images obtained from
the {\sl HST} Data Archive were observed through the F343N and F438W
passbands (program ID: GO-15062, PI: N. Bastian), while the WFPC2
images provide the corresponding observations in the F555W and F814W
passbands (program ID: GO-8141, PI: R. M. Rich). The UVIS/WFC3 data
set is composed of three frames taken through the F343N passband, with
exposure times of 540 s and 1060 s (twice), as well as three frames
taken through the F438W passband, with exposure times of 120 s and 550
s (twice). For both the F555W and F814W passbands, the WFPC2 data set
contains four frames each, for each frame, the exposure time is 400 s. 

Similarly to our previous papers \cite[e.g.,][]{Li17a}, we applied
point-spread-function (PSF) photometry to the `{\tt \_flt}' and `{\tt
  \_c0f}' frames based on the standard recipes recommended by the {\sc
  Dolphot2.0} package \citep{Dolp11a,Dolp11b,Dolp13a}. { {\sc
    Dolphot2.0} is a photometric package specifically designed for
  {\sl HST} photometric analysis. We use its WFC3 and WFPC2 modules to
  deal with the relevant observational data. They include built-in
  charge-transfer efficiency corrections and photometric calibration
  routines such as aperture and zeropoint corrections. {\sc
    Dolphot2.0} has been validated extensively; it is one of the most
  powerful tools for {\sl HST} photometric analyses
  \citep[e.g.,][]{Mone10a}}.
  
In our photometry, only objects meeting the following criteria were
selected as `good' stars: (1) flagged by {\sc Dolphot2.0} as a `bright
star.' (2) Not centrally saturated. (3) Sharpness is between $-$0.3
and 0.3. (4) Crowding parameter $< 0.5$. Our photometric approach
resulted in identification of 19,507 and 10,915 stars in the UVIS/WFC3
and WFPC2 frames, respectively. We carefully combined both output
catalogs by cross-matching the stars in common. Our final, combined
stellar catalog contains 6856 stars.

\section{Main Results}

\begin{figure*}
\includegraphics[width=2\columnwidth]{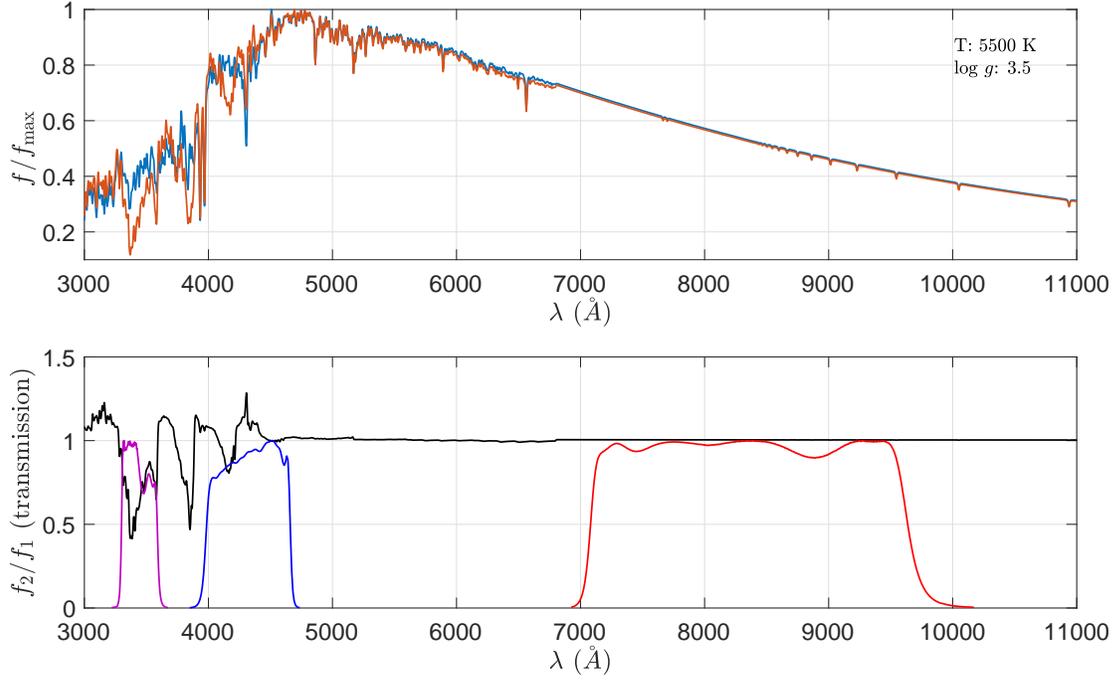}
\caption{ (top) Model spectra for different CNO abundances. The
  blue spectrum represents stars with `normal' abundances, while the
  red spectrum represents stars with enhanced nitrogen and depleted
  carbon and oxygen abundances ($\Delta$[N/Fe] = 1.0 dex,
  $\Delta$[C/Fe] = $\Delta$[O/Fe] = $-$0.6 dex). (bottom)
  Corresponding flux ratio and normalized filter transmission curves
  used here (from left to right: F343N/WFC3, F438W/WFC3,
  F814W/WFPC2). The adopted model metallicity is [Fe/H] = $-$0.65 dex,
  as determined through isochrone fitting. The adopted temperature and
  log $g$ values pertain to stars at the base of the RGB. Our model
  spectra were smoothed with a Gaussian kernel defined by $\sigma$ =
  10$\AA$.}
\label{F1}
\end{figure*}

The most salient feature of the MPs in most GCs and intermediate-age
clusters is the star-to-star variations in light elements (e.g., C, N,
O, Na). Stars in these clusters usually have different C, N, and He
abundances, and variations in these elements could broaden or split
the clusters' RGBs. Specifically, an N spread would strengthen the NH
molecular feature at $\sim3370${\AA}, thus causing a variable RGB
morphology in CMDs partially defined by UV filters (specifically,
F343N). In addition, the {\sl HST}'s F438W passband includes the CH
absorption feature ($\sim4300${\AA}). Pristine and enriched stellar
populations will exhibit different F343N$-$F438W colors owing to their
different N (and, to a lesser extent, C) abundances. N-rich stars will
appear redder than N-poor stars. On the other hand, since N-rich stars
should, in principle, also be He-rich, they are generally hotter than
He-poor stars of the same luminosity. Helium variations among cluster
member stars could thus also be revealed by examination of their
F438W$-$F814W colors. { This is illustrated in Figure \ref{F1}: in
  the top panel we show two model spectra for stars at the base of the
  RGB, with normal and enriched nitrogen abundances ([Fe/H]=$-$0.65
  dex, $\Delta$[N/Fe]=1.0 dex). For the model characterized by an
  enriched nitrogen abundance, we also reduced its carbon and oxygen
  abundances so that the total CNO abundance does not vary (i.e.,
  $\Delta$[C+N+O/Fe]=0 dex). Our model spectra were calculated using
  the {\sc Spectrum 2.77} package
  \citep{Gray94a}\footnote{\url{http://www.appstate.edu/~grayro/spectrum/spectrum.html}}
  and based on the ATLAS9 stellar atmosphere models
  \citep{Kuru70a,Kuru93a}}

\begin{figure*}
\includegraphics[width=2\columnwidth]{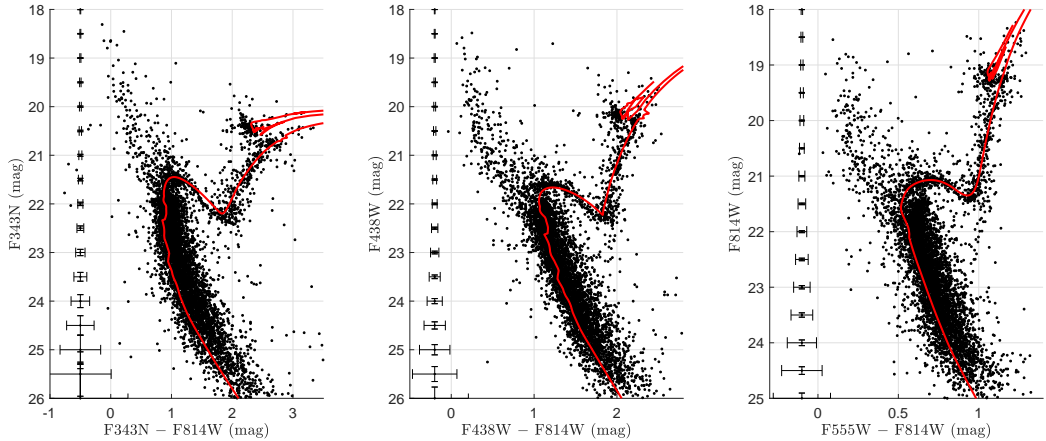}
\caption{NGC 121 CMDs. (left) F343N$-$F814W vs F343N; (middle)
  F438W$-$F814W vs F438W; (right) F555W$-$F814W vs F555W. The red
  lines are the best-fitting isochrones. In each panel, the errorbar is on the left side.}
\label{F2}
\end{figure*}

We note that the RGB of NGC 2121 is severely contaminated by a young
field-star population in CMDs involving the F343N and F438W
filters. This problem can only be ameliorated by introducing
additional observations in the F814W passband (i.e., the WFPC2
observations). Figure \ref{F2} shows three CMDs involving the F343N,
F438W, and F555W passbands. At first glance, we find that although the
cluster's RGB is tight in the F438W--F814W vs F438W (middle) and
F555W--F814W vs F555W (right) CMDs, it shows a moderate broadening in
the F343N--F814W vs F343N CMD (left).

To quantify any broadening of the RGB caused by chemical variations,
we simulated multi-band photometry of our target cluster. We first
used the MESA Isochrone and Stellar Tracks
\citep[MIST;][]{Paxt11a,Paxt13a,Paxt15a,Choi16a,Dott16a} models to
generate the best-fitting isochrones representative of the
observations. { Recently, \cite{Bark18a} reported a problem
  associated with isochrone fitting to {\sl HST} photometry in
  UV--optical--IR passbands. This problem has also been recognized by
  the community at large
  \citep[e.g.,][]{Gont19a,Howe19a}. Specifically, model parameters
  determined through isochrone fitting to the optical--IR CMD cannot
  be used to adequately describe photometric measurements involving UV
  passbands. In this paper, we encountered the same problem. We found
  that we were unable to identify a set of model parameters that allow
  us to simultaneously fit all three CMDs in Figure \ref{F2}. Our
  best-fitting age, metallicity, and distance modulus were determined
  based on the CMD involving the F555W and F814W passbands, resulting
  in $\log{(t\;{\rm yr}^{-1})} =9.51\pm0.02$, [Fe/H] = $-0.65\pm0.10$
  dex, and $(m-M)_0 =18.42\pm0.05$ mag. The associated uncertainties
  were determined by the size of grids used for the fit. We determined
  the best-fitting age by visually inspecting the fit to the SGB. The
  metallicity was determined by fitting the slope of the cluster's
  RGB. The best-fitting distance modulus was determined by fitting the
  magnitude of the red clump. We found that once we had fixed the
  best-fitting extinction based on one of our diagnostic CMDs, the
  relevant isochrone would exhibit a moderate offset in UV
  passbands. We would have to adopt different extinction values for
  each CMD to obtain best fits, specifically $E(B-V)$ = 0.09, 0.10,
  and 0.12 mag (for the CMDs from the left to the right in Figure
  \ref{F2}). Since this is unphysical, it is more likely that these
  offsets may have been caused by some unknown calibration limitation
  of the MIST models across different photometric systems. { We
    agree with \cite{Bark18a}, who warned that ``until the models
    are fixed, they should not be used for fitting or determining
    stellar populations in the UV.'' } { Therefore, in this paper
    we only focus on the width of the cluster's RGB, which should not
    be affected by the model's limitations. We also suggest that the
    extinction derived from the optical--IR CMD likely represents the
    most accurate value, $E(B-V)=$ 0.12 mag.}}

Our best-fitting age and metallicity are consistent with the values
obtained by \cite{Rich01a}. However, previously determined extinction
values vary from $E(B-V) =0.07$ mag to 0.14 mag
\citep{Udal98a,Kerb07a}. Our newly derived distance modulus to NGC
2121 is slightly smaller than the canonical LMC value \citep[$(m-M)_0
  =18.50$ mag; e.g.,][]{grijs14a}. We visually confirmed that our fits
adequately describe most of the CMD sequences. The best-fitting
isochrones to each of the CMDs are presented in Figure \ref{F2}.

To quantify whether NGC 2121 has a broadened RGB caused by a chemical
spread, we need to compare the observed CMD with that of a simulated
SSP. In principle, in addition to chemical spreads, several other
factors may also cause a broadening of the RGB, including (i)
photometric uncertainties, (ii) photometric artefacts (cosmic rays,
bad or hot pixels, etc.), (iii) differences in distances to individual
stars, (iv) differential reddening, and (v) field-star
contamination. Any broadening caused by distance differences to the
cluster stars is negligible because of the large distance to the
LMC. Photometric uncertainties and artefacts can be assessed based on
artificial-star tests. For the images observed with each camera, we
generated 28,000 artificial stars located on the best-fitting
isochrone between the onset of the SGB and the upper part of the
RGB. Their spatial distributions were homogeneous. To avoid a
situation where artificial stars dominate the background and crowding
levels, we only added 100 artificial stars to the raw images at any
one time. We used the same photometric method to measure these input
stars and applied the same data reduction as used for the observations
to the artificial stellar catalog. Finally, we recovered 24,943
artificial stars from the WFPC2 frames, corresponding to a
completeness level of 89\% for the SGB and RGB stars. From the
UVIS/WFC3 observations, we recovered 27,792 artificial stars,
indicating a completeness level of close to 99\%.

\begin{figure}
\includegraphics[width=\columnwidth]{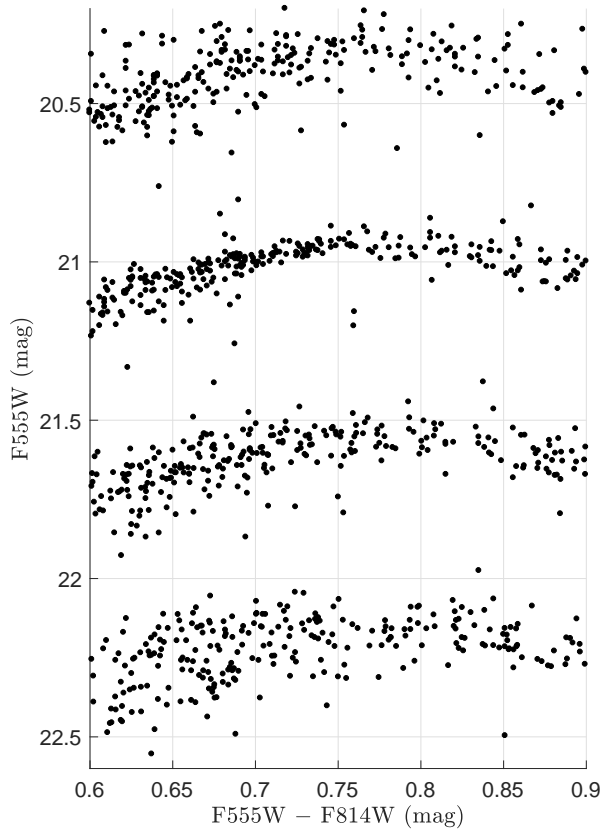}
\caption{Observed SGB of NGC 2121 (top) compared with simulated SGBs
  characterized by different degrees of differential reddening. From
  the second to the bottom, the amounts of reddening adopted
  correspond to $\Delta{A_V}= 0.0, 0.04$, and 0.08 mag.}
\label{F3}
\end{figure}

The recovered artificial stellar population should be affected by the
same photometric uncertainties and artefacts as the
observations. However, it cannot reveal the level of internal
differential reddening. \cite{Milo12a} developed a statistical method
to study the reddening distribution in small areas (such as the core
regions of GCs). However, this method is not applicable to our target
cluster, because the number of stars is small \citep[i.e., less than
  10\% of the numbers observed for most GCs; see
][]{Milo12a}). Fortunately, NGC 2121 exhibits a tight SGB. In the CMD
based on the F555W and F814W filters, any broadening caused by
chemical inhomogeneities is negligible. Therefore, any additional
broadening that cannot be reproduced by photometric uncertainties or
artefacts must be caused by differential reddening. As such, we
compared the width of the simulated SGB to the observations and
determined the best-fitting differential reddening level. We conclude
that the degree of differential reddening in NGC 2121 is likely of
order $\Delta{A_V}=0.04\pm0.01$ ($\Delta{E(B-V)}=0.013\pm0.003$): see
Figure \ref{F3}

\begin{figure}
\includegraphics[width=\columnwidth]{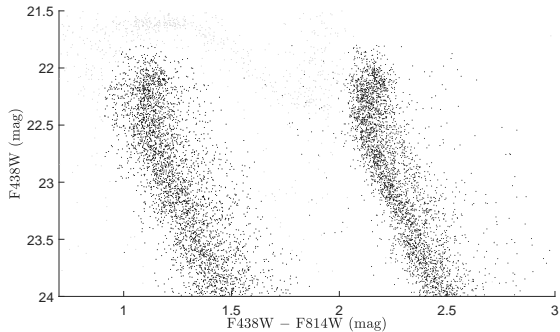}
\caption{ Observed MS and MSTO region of NGC 2121 (left) compared
  with the model MS and MSTO including photometric uncertainties and
  differential reddening (right).}
\label{F4}
\end{figure}

{ Using artificial stars, we can also explore if NGC 2121 may host
  an eMSTO similar to its younger counterparts \citep{Milo09a}. We
  generated 20,000 artificial stars based on the observational MS
  ridge-line. We added these artificial stars to the observational
  images and measured them using the same photometric method as used
  for the real observations. We also applied differential reddening to
  these artificial stars. We found that the combination of photometric
  uncertainties and artefacts, combined with the differential
  reddening, is sufficient to reproduce the observed width of the MSTO
  region. Therefore we conclude that NGC 2121 does not feature an
  eMSTO. In addition, its apparently tight SGB, shown in Figure
  \ref{F3}, is also known to represent a coeval stellar population
  \citep[e.g.,][]{Li14a,Li16a}. In Figure \ref{F4} we show an example
  comparison between the observations and the simulation in the
  F438W$-$F814W vs F438W CMD.}

\begin{figure}
\includegraphics[width=\columnwidth]{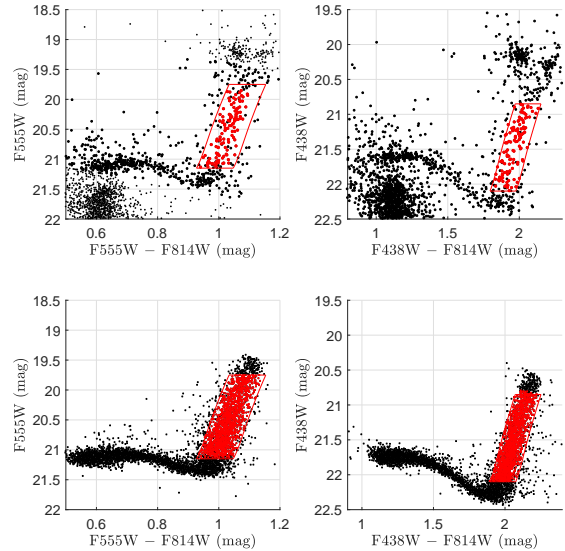}
\caption{Selection of RGB stars from the observations. Only stars
  located in both selection boxes are considered cluster member
  stars. The selection boxes were defined based on the simulated CMDs
  (bottom).}
\label{F5}
\end{figure}

Because of the large distance to the LMC, using stellar proper motions
to reduce field contamination is not possible. In addition, the field
of view of the combined observations is very small, rendering
selection of an appropriate field region for reference purposes
troublesome. To minimize the impact of field-star contamination, we
selected RGB stars according to their distribution in two of our CMDs
(i.e., F555W--F814W vs F555W and F438W--F814W vs F438W). We did not
select RGB stars from the CMD involving F343N photometry, because in
that section of parameter space RGB stars may have been affected by
star-to-star variations in N. We used the simulated CMDs to determine
the typical regions occupied by the majority of RGB stars and
subsequently used those regions to select RGB stars from our
observational parameter space: see Figure \ref{F5}. Only stars located
in the RGB selection boxes in both CMDs were considered RGB stars. As
shown by \cite{Mart17a}, selecting RGB stars from multiple CMDs can be
used to effectively reduce field-star contamination.

\begin{figure*}
\includegraphics[width=2\columnwidth]{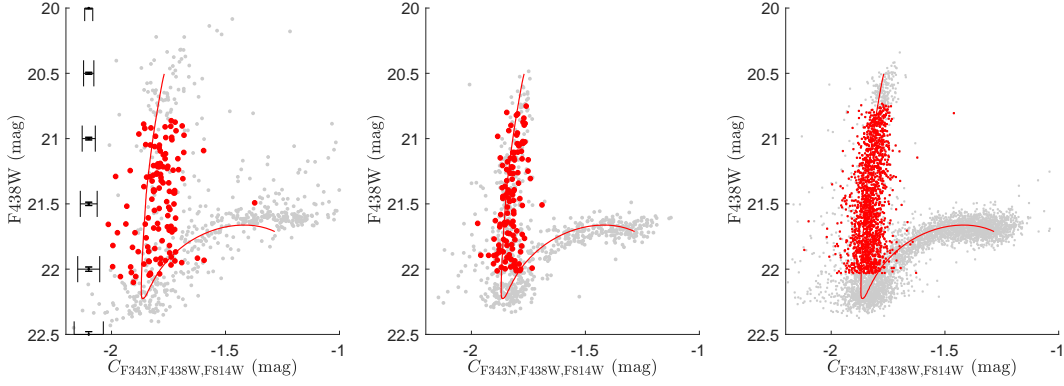}
\caption{F438W vs $C_{\rm F343N,F438W,F814W}$ diagram for the
  observations (left panel, with the errorbar on the left side) and the simulations (middle and right). The middle panel shows a simulated subsample (with the same number of
  stars as the observations), while the right-hand panel includes all
  simulated stars.}
\label{F6}
\end{figure*}

\begin{figure}
\includegraphics[width=\columnwidth]{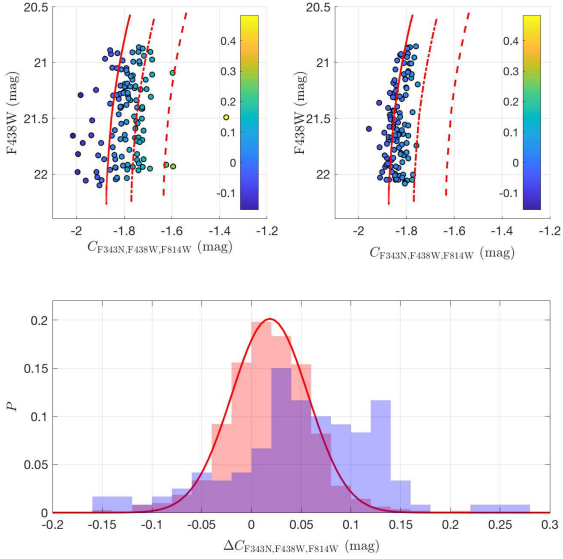}
\caption{F438W vs $C_{\rm F343N,F438W,F814W}$ diagram for the
  observations (top left) and our simulations (top right), with their
  $\Delta(C_{\rm F343N,F438W,F814W})$ color-coded. The solid, dashed,
  and dash-dotted lines indicate the best-fitting isochrone
  ($\Delta$[C/Fe] = $\Delta$[N/Fe] = $\Delta$[O/Fe] = 0.00 dex) and
  the loci for $\Delta$[N/Fe] = 0.50 dex ($\Delta$[C/Fe] =
  $\Delta$[O/Fe] = $-$0.09 dex) and $\Delta$[N/Fe] = 1.00 dex
  ($\Delta$[C/Fe] = $\Delta$[O/Fe] = $-$0.60 dex). The bottom panel
  shows the $\Delta(C_{\rm F343N,F438W,F814W})$ probability
  distributions for the observations (red histogram) and the
  simulation (blue histogram). The distribution of $\Delta(C_{\rm
    F343N,F438W,F814W})$ for the simulated SSP RGB can be well
  described by a singe Gaussian profile (red curve).}
\label{F7}
\end{figure}

We used a similar method as \cite{Mone13a} to quantify any broadening
of the RGB caused by MPs. We constructed the color index $C_{\rm
  F343N,F438W,F814W}$ = (F343N$-$F438W)$-$(F438W$-$F814W). This
pseudo-color is an effective index to uncover MPs
\citep{Mone13a,Mart17a}. We next compared the observational and
simulated (artificial) distributions of stars in the F438W versus
$C_{\rm F343N,F438W,F814W}$ diagrams: see Figure \ref{F6}. The
simulated diagram on the right includes all 25,000 stars. We also
randomly selected a subsample of our artificial stars composed of the
same number of stars as in our observations, i.e., only containing 120
RGB stars (middle panel; red circles). Figure \ref{F6} shows that the
observed RGB exhibits a larger spread than the simulated SSP RGB. To
quantify this broadening, we adopted the best-fitting isochrone as our
fiducial line and calculated the corresponding deviations of the
pseudo-color indices, as illustrated in Figure \ref{F7}. The
distributions of the pseudo-color indices of the observed and
simulated RGB stars are presented in Figure \ref{F7} as well.

\begin{figure}
\includegraphics[width=\columnwidth]{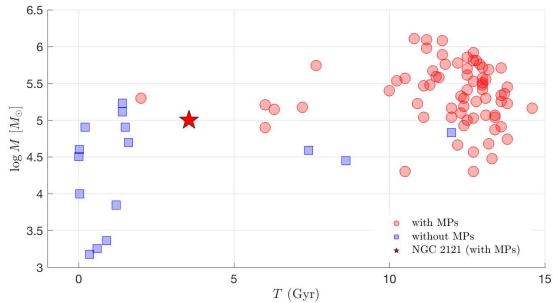}
\caption{Age--mass plane for young and intermediate-age clusters and
  GCs with and without MPs; data have been derived from
  \cite{Mcla05a,Baum13a,Krau16a}. Blue squares are clusters without
  MPs (or where the presence of MPs is unclear). Red circles are
  clusters with MPs. The red pentagram indicates the locus of NGC
  2121.}
\label{F8}
\end{figure}

From inspection of Figure \ref{F7}, { we found that the peak of the
  observed distribution of the pseudo-colors for the simulated SSP is
  not consistent with the observations. Again, this is likely owing to
  the fitting problem encountered when UV observations are
  involved. Therefore, we only focus on the internal spread of the
  pseudo-color distribution rather than the actual pseudo-color
  values}. We found that the distribution of the pseudo-colors for the
simulated SSP can be adequately described by a single Gaussian
profile,
\begin{equation}
P(\Delta{C})=0.201e^{-\left({\frac{\Delta{C}-0.019}{0.054}}\right)^2},
\end{equation}
with a standard deviation of $\sigma=0.054$ mag. The observed
pseudo-color distribution of the red-giant stars, however, is not
well-described by a single Gaussian distribution; it is much broader
than the distribution resulting from the simulation. If we were to
force a single Gaussian profile to fit the distribution, the
`best-fitting' function would be
\begin{equation}
P(\Delta{C})=0.118e^{-\left({\frac{\Delta{C}-0.061}{0.090}}\right)^2}.
\end{equation}
The corresponding standard deviation would be $\sigma=0.090$ mag,
about five-thirds that of the simulated SSP. To examine if the more
broadened pseudo-color distribution of the observed red-giant stars
could simply have been caused by small-number statistics, we selected
10 subsamples from the simulated stars containing the same numbers of
stars as the observation. All subsamples exhibit more dispersed
pseudo-color distributions than the observations (Figure \ref{F8}).

{ Using ATLAS9 atmosphere models \citep{Kuru70a,Kuru93a}, we
  calculated 16 model spectra with atmosphere parameters ([Fe/H], $T$,
  $\log$ $g$) that were the same as those adopted by the best-fitting
  isochrones. These model spectra represent stars between the base and
  the middle of the RGB, i.e., they are for `standard' stars with
  `normal' abundances (i.e., $\Delta$[C/Fe] = $\Delta$[N/Fe] =
  $\Delta$[O/Fe] = 0.00 dex). We then calculated a set of N-enhanced
  models with $\Delta$[N/Fe] = 0.50 dex, $\Delta$[C/Fe] =
  $\Delta$[O/Fe] = $-$0.09 dex, and $\Delta$[N/Fe] = 1.00 dex,
  $\Delta$[C/Fe] = $\Delta$[O/Fe] = $-$0.60 dex. Thus, we calculated
  three sets of model spectra with no, modest, and strong N
  enrichment. Again, we fixed the total CNO abundance by reference to
  that observed in GCs \citep[e.g.,][]{Mari16a}. We calculated the
  flux ratio compared with its normal counterpart for every enriched
  star by folding their model spectra through the corresponding filter
  transmission curve. We converted the flux ratios in all passbands
  into magnitude differences. Finally, we obtained an isochrone
  characterized by suitable CNO abundances anomalies. As shown in the
  top panels of Figure \ref{F7}, the loci for different CNO abundances
  are roughly parallel from the bottom to the middle of the RGB. This
  result is similar to that of \cite{Mart17a}. Our fits show that the
  observed width of the RGB in the F438W versus $C_{\rm
    F343N,F438W,F814W}$ diagram can be explained by invoking chemical
  variations from standard abundances up to $\Delta$[N/Fe] = 1.0 dex
  (with $\Delta$[C/Fe] = $\Delta$[O/Fe] = $-$0.6 dex). If we were to
  force the standard isochrone to meet the peak of the pseudo-color
  distribution, an internal abundance spread of $\Delta$[N/Fe] = 0.5
  dex would still be required. We suggest that the actual N abundance
  spread among the RGB stars in NGC 2121 might be in the range
  0.5--1.0 dex.} Since the observed pseudo-color dispersion of the
red-giant stars cannot be explained by photometric uncertainties,
artefacts, differential reddening, or small-number statistics, the
only viable explanation is light-element star-to-star variations,
i.e., the RGB of NGC 2121 appears to be composed of MPs.

In Figure \ref{F8} we have added NGC 2121 to the age--mass plane for
clusters with and without MPs, which represents a summary of results
from the literature as to whether MPs occur in GCs and their younger
massive counterparts of different ages and masses. Almost all old GCs
show MPs, while their younger counterparts (younger than 2 Gyr) do
not. In the age range of 2--10 Gyr, clusters with and without MPs
overlap in both age and mass. The mass of NGC 2121, $\sim10^5$
$M_{\odot}$ \citep{Mcla05a}, is only half that of NGC 1978
\citep[$\sim2\times10^5$ $M_{\odot}$;][]{Baum13a}, and its mass is
comparable to most of its younger counterparts.

\section{Discussion and Summary}

In this paper, we have analyzed the photometric appearance of the NGC
2121 RGB in the F438W versus $C_{\rm F343N,F438W,F814W}$ diagram,
which has been established as a diagnostic plot suitable for
uncovering the presence of MPs. A broadening of the cluster's RGB is
apparent when compared with that of a simulated SSP. It is consistent
with a model characterized by different CNO abundances, implying
star-to-star chemical variations among the cluster's red-giant stars.

\begin{figure*}
\includegraphics[width=2\columnwidth]{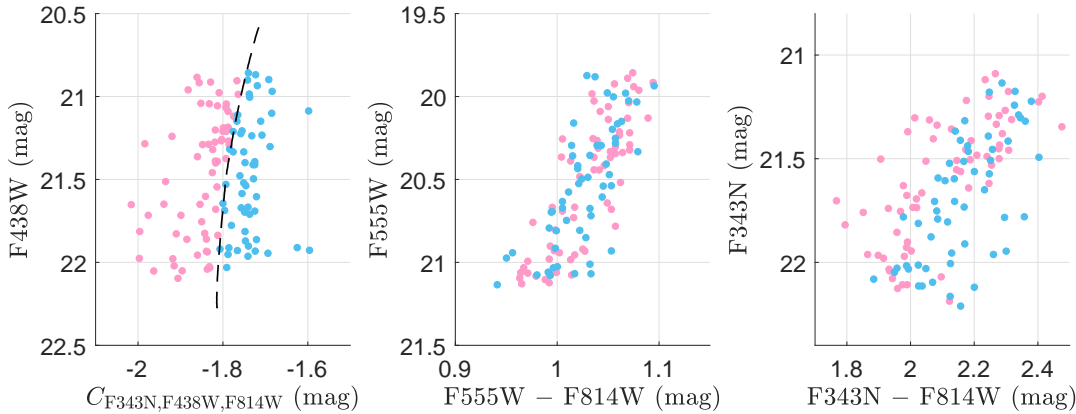}
\caption{NGC 2121 RGB in the F438W vs $C_{\rm F343N,F814W,F814W}$
  diagram (left). Stars are divided into two subsamples based on the
  dividing line halfway between the standard isochrone and the locus
  for $\Delta$[N/Fe] = 0.5 dex (see the pink and blue circles). The
  middle and right-hand panels present the same red-giant stars in the
  other CMDs.}
\label{F9}
\end{figure*}

We have determined that F343N is indeed a key passband for use to
unveil the presence of different stellar populations. We divided our
selected RGB stars into two subsamples based on the dividing line
halfway between the standard isochrone and the locus pertaining to
$\Delta$[N/Fe] = 0.5 dex. We explored their color distributions in
other CMDs, as shown in Figure \ref{F9}. Indeed, both subsamples are
fully mixed in the F555W$-$F814W vs F555W CMD, but they exhibit
distinct color differences in the CMD involving the F343N and F814W
passbands. Stars with larger $C_{\rm F343N,F438W,F814W}$ indices have
redder F343N$-$F814W colors, indicating that their total fluxes in the
F343N passband are much lower than those of their blue counterparts,
i.e., they are N-enriched stars. Our detection of chemical variations
among the red-giant stars in NGC 2121 also makes it the
second-youngest cluster with MPs (after the 2.2 Gyr-old cluster NGC
1978).

It would be interesting to examine whether the pristine and enriched
stellar populations in the cluster have different central
concentrations. However, the spatial distribution of stars across the
WF3 chip of the WFPC2 camera exhibits numerous areas with little or no
stars, indicating that source confusion caused by the spikes
associated with bright stars in this region is severe. Therefore, the
corresponding stellar completeness varies significantly at different
radii. High-quality WFC3 observations of a larger field of view,
through the F814W passband, are required to resolve this problem.

It is useful to compare the physical properties of NGC 2121 with those
of its younger counterparts without MPs, i.e., NGC 1783, NGC 1806, and
NGC 1846 \citep{Mucc14a,Mart18a,Zhang18a}. The masses of NGC 1783, NGC
1806, NGC 1846, and NGC 2121 are all similar
\citep[$\sim1.8\times10^5$ $M_{\odot}$, $10^5$ $M_{\odot}$,
  $1.2\times10^5$ $M_{\odot}$, and $10^5$ $M_{\odot}$,
  respectively;][]{Mcla05a,Baum13a}, as are their internal structural
parameters \citep[such as their core, half-mass, and tidal
  radii;][]{Mcla05a,Li18a}. These clusters also have similar
metallicities and distances to the LMC's bar region
\citep{Li18a}. Thus, both the external environments and the internal
dynamical properties of NGC 2121 and its younger counterparts are
similar, which implies that the presence of MPs in NGC 2121 is
unlikely caused by any specific formation environment or internal
dynamics.

A noticeable difference with NGC 1783, NGC 1806, and NGC 1846 is that
these younger clusters exhibit apparent eMSTO regions while NGC 2121
and NGC 1978 \cite{Mart18a} do not. \cite{Mart18a} proposed that the
apparent chemical anomalies might be a specific feature of stars with
masses below 1.5 $M_{\odot}$, which is roughly the mass of
main-sequence turnoff stars at an age of $\sim$2 Gyr. Stars with
masses below this critical mass would be affected by magnetic braking,
thus making them all slow rotators. Only stars with masses greater
than 1.5 $M_{\odot}$ will exhibit evidence of rapid rotation in their
cluster CMD, as manifested by eMSTO regions. Although the details are
still unclear, strong magnetic fields may play a role in the
appearance of star-to-star chemical variations. This notion is
supported by our results, since the masses of the RGB and MSTO stars
in NGC 2121 have decreased to below this critical mass. All of these
stars should possess strong magnetic fields.

The detection of MPs in NGC 2121 underpins the hypothesis that age may
be an important factor controlling the presence of MPs. The transition
period between clusters with and without MPs should occur at an age of
2--3 Gyr. The mass of NGC 2121 is comparable to those of most Galactic
GCs exhibiting MPs. It is unclear whether cluster mass also plays a
role in the appearance of MPs. However, if so, it should not be the
only factor of importance, because otherwise the detection of MPs in
younger massive clusters should also be expected. This conclusion is
consistent with that of \cite{Zhang18a}. It is important to search for
MPs in other clusters of similar ages but lower masses than NGC 2121,
such as NGC 2193 and ESO-56-SC40 \citep{Baum13a}. If these clusters
exhibit similar CMD features as NGC 1978 and NGC 2121, this would lead
to the conclusion that mass is not a crucial parameter determining the
presence of MPs. Future studies should then focus on which intrinsic
transitions may have led to chemical star-to-star variations at ages
between 2 and 3 Gyr. On the other hand, if such clusters do not
exhibit MPs, this would imply that mass may still be a secondary
parameter controlling the presence of MPs.

\acknowledgements 
C. L. was supported by the Macquarie Research Fellowship Scheme. This
work was also partly supported by the National Natural Science
Foundation of China through grants U1631102, 11373010, 11633005, and
11803048.


\facilities{{\sl Hubble Space Telescope} (UVIS/WFC3 and WFPC2)}
\software{\sc dolphot2.0 \citep{Dolp11a,Dolp11b,Dolp13a}}
\\{\sc Spectrum v2.77 \citep{Gray94a}}

\end{document}